\documentclass[twocolumn,showpacs,superscriptaddress,floatfix,preprintnumbers,amssymb,amsmath]{revtex4}

\usepackage{graphicx}
\usepackage{dcolumn}
\usepackage[latin1]{inputenc}
\usepackage[mathscr]{eucal}
\usepackage{epsfig}
\usepackage{rotating}
\usepackage{bm}

\begin{document}
\preprint{ }

\title{Vortex pump for dilute Bose-Einstein condensates}

\author{Mikko M\"ott\"onen}
\affiliation{Laboratory of Physics, Helsinki University of Technology P. O. Box 4100, FI-02015 TKK, Finland}
\affiliation{Low Temperature
Laboratory, Helsinki University of Technology, P.O. Box 3500, 02015 TKK, Finland}
\author{Ville Pietil\"a}
\affiliation{Laboratory of Physics, Helsinki University of Technology P. O. Box 4100, FI-02015 TKK, Finland}
\author{Sami M.\ M.\ Virtanen}
\affiliation{Laboratory of Physics, Helsinki University of Technology P. O. Box 4100, FI-02015 TKK, Finland}

\begin{abstract}
The formation of vortices by topological phase engineering has
been realized experimentally to create the first two- and
four-quantum vortices in dilute atomic Bose-Einstein
condensates~[A.\ E.\ Leanhardt {\it et al.}, Phys. Rev. Lett. {\bf
89}, 190403 (2002)]. We consider a similar system, but in addition
to the Ioffe-Pritchard magnetic trap we employ an additional
hexapole field. By controlling cyclically the strengths of these
magnetic fields, we show that a fixed amount of vorticity can be
added to the condensate in each cycle. In an adiabatic operation
of this vortex pump, the appearance of vortices into the
condensate is interpreted as the accumulation of a local Berry
phase. Our design can be used as an experimentally realizable
vortex source for possible vortex-based applications of dilute
Bose-Einstein condensates.
\end{abstract}
\maketitle

Quantized vortices manifest the long-range phase coherence of
many-particle quantum systems which are described by a
complex-valued order parameter field, and their existence and
stability is intimately related to the superfluid properties of
the system~\cite{Pethick2002a}. In a loop encircling the vortex,
the phase of the order parameter undergoes an integer
multiple~$\kappa$ of $2\pi$ winding. This number~$\kappa$ is the
circulation quantum number of the vortex.

The first vortices in trapped gaseous Bose-Einstein condensates
(BECs) of atoms were created by driving dynamically a transition
between two different hyperfine spin states of the condensed atoms
with a rotating laser field~\cite{Matthews1999b}, as suggested in
Ref.~\cite{Williams1999c}. Single-quantum vortices have also been
nucleated by slicing through the condensate with a focused laser
beam moving faster than the critical velocity~\cite{Inouye2001a},
by colliding condensates separated by tailored optical
potentials~\cite{Scherer2007}, by stirring the condensate with a
laser beam~\cite{Madison2000a}, and by rotating the condensate
with an asymmetric trapping potential~\cite{Hodby2001}. In
harmonic traps, multi-quantum vortices are unstable against
splitting into single-quantum
vortices~\cite{Mottonen2003a,Shin2004a,Huhtamaki2006a}, and hence
rotation of the condensate with frequencies close to the harmonic
trapping frequency results in triangular vortex lattices of
single-quantum vortices~\cite{Raman2001a,Abo-Shaeer2001a}.
Recently, multi-quantum vortices have been created by a coherent
transfer of angular momentum from a photon of a Laguerre-Gaussian
laser beam to the condensate~\cite{Andersen2006}. Furthermore,
giant vortices with circulations up to 60 quanta have been created
from such vortex lattices with a tightly focused laser removing
atoms from the center of the condensate for the giant vortex to
appear~\cite{Engels2003}.

In contrast to the above-mentioned dynamical methods,
vortices can also be created
with the so-called topological phase engineering
technique~\cite{Nakahara2000a,Isoshima2000a,Mottonen2002a,Mottonen2002b}.
Here, the phase winding of the vortex is generated by an
azimuthally dependent adiabatic turn of the hyperfine spin of the
condensate. The first two- and four-quantum vortices were created
in dilute BECs with this method by Leanhardt et
al.~\cite{Leanhardt2002a}. They reversed the bias field of the
Ioffe-Pritchard trap and verified that the resulting two- and
four-quantum vortex states were consistent with the angular
momenta of~$2\hbar$ and~$4\hbar$ per particle, respectively. After
the reversal, the bias magnetic field was brought back to its
initial configuration and the topological unwinding of the
vortices was observed.

In this Letter, we show that by controlling the strengths of the
bias and the quadrupole fields of the Ioffe-Pritchard trap and an
additional hexapole magnetic field, vortex pumping can be
implemented, i.e., the quantum number of the vortex in the
condensate is increased by a constant amount in each pumping
cycle. The pumping can be carried out either fully adiabatically
or partly adiabatically and partly instantaneously. The operation
of the pump is studied analytically within the adiabatic
approximation, and computationally using the Gross-Pitaevskii (GP)
equation for hyperfine spin $F=1$ condensates. In this special
case, the reversal of the bias field in the presence of the
hexapole field creates a four-quantum vortex. When the bias field
is brought back to its initial configuration in the presence of
the quadrupole field, two quanta of vorticity is lost. Thus the
pump creates two net quanta of vorticity per cycle. To date, the
vortex pump introduced here is the only known method to produce
almost any desired amount of vorticity without particle loss.
Furthermore, the achieved vortex densities can possibly be high
enough for the experimental realisation of strongly correlated
vortex liquid phases~\cite{Cooper2001}.

In the adiabatic approximation for an atom with a positive
Land{$\acute{\textrm{e}}$}~$g$ factor, the spin of the condensate
is antiparallel to the local magnetic field
$\bm{B}=\bm{B}_\perp+\bm{B}_z$, where~$\bm{B}_z$ is the uniform
bias field along the positive $z$-direction and the perpendicular
part~$\bm{B}_\perp$ is either the quadrupole or the hexapole field
given in the polar coordinates by
\begin{eqnarray}\label{Bq}
\bm{B}_q(r,\phi,t)&=&B_q'(t)r[\cos(\phi)\hat{\bm{x}}-\sin(\phi)\hat{\bm{y}}],
\\ \label{Bh}
\bm{B}_h(r,\phi,t)&=&B_h'(t)r[\cos(2\phi)\hat{\bm{x}}-\sin(2\phi)\hat{\bm{y}}],
\end{eqnarray}
respectively. Here,~$\hat{\bm{x}}$ and~$\hat{\bm{y}}$ are the unit
vectors along the positive~$x$- and~$y$-directions, respectively.
The above fields up to quadratic terms in the coordinates can be
realized experimentally with five coils in total and the field
strengths~$B_q'(t)$ and~$B_h'(t)$ can be controlled by changing
the currents in the coils~\cite{Pethick2002a}.

Let us assume that the adiabatic approximation is valid and the
bias field~$\bm{B}_z$ is initially dominating over the
perpendicular field such that the spin of the condensate is
aligned with the $z$-axis, i.e., is in the weak-field-seeking
state $|F,m_z=-F\rangle$, where~$m_z$ is the quantum number of the
spin along the $z$-direction. Thus the reversal of the bias field
introduces a spin rotation about the normal vector~$\hat{\bm{n}}$
of the perpendicular field in the $xy$-plane as
\begin{equation}\label{rot}
R_{\hat{\bm{n}}}(\pi)=e^{-i\pi\bm{\mathcal{F}}\cdot\hat{\bm{n}}},
\end{equation}
where the dot product is defined as $\bm{\mathcal{F}}\cdot\hat{\bm{n}}=\mathcal{F}_xn_x+\mathcal{F}_yn_y+\mathcal{F}_zn_z$ and
$\mathcal{F}_\alpha$ is the dimensionless spin operator in the direction~$\alpha$. 
From Eqs.~\eqref{Bq} and~\eqref{Bh} we obtain the normal vectors
\begin{eqnarray}\label{nq}
\hat{\bm{n}}_q(\phi)&=&\sin(\phi)\hat{\bm{x}}+\cos(\phi)\hat{\bm{y}},\\
\label{nh}
\hat{\bm{n}}_h(\phi)&=&\sin(2\phi)\hat{\bm{x}}+\cos(2\phi)\hat{\bm{y}},
\end{eqnarray}
for the quadrupole and hexapole fields, respectively. Thus the
reversal of the bias field in the presence of the hexapole field
steers the initial spin state to
\begin{equation}\label{vf}
R_{\hat{\bm{n}}_h}(\pi)|F,m_z\rangle=(-1)^{F+m_z}e^{-i4m_z\phi}|F,-m_z\rangle,
\end{equation}
changing the vorticity of the system by~$-4m_z$. The above result
can be obtained by expressing the rotation with the Euler angles
as
$R_{\hat{\bm{n}}_h}(\pi)=R_{\hat{\bm{z}}}(\pi/2-2\phi)R_{\hat{\bm{x}}}(\pi)R_{\hat{\bm{z}}}(2\phi-\pi/2)$.

\begin{figure}[h]
    \begin{center}
   \includegraphics[width=.45\textwidth]{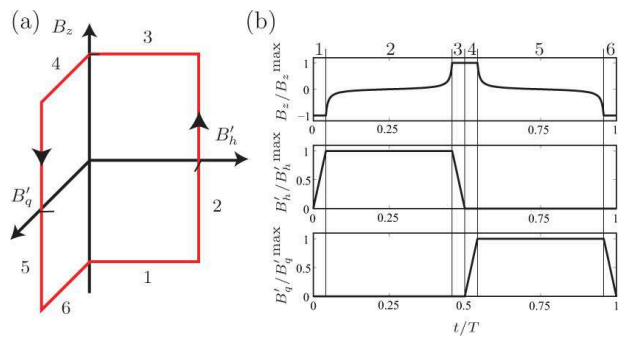}
    \end{center}
   \caption{\label{fig1} (color online). Possible control cycle in the magnetic field parameter space for pumping vortices into a spinor Bose-Einstein condensate (a). Panel (b) displays an example of the temporal changes of the control fields such that the spin at a certain distance from the $z$-axis is reversed with a constant speed. The different parts of the control loop in panel (a) are marked in the top of panel (b).}
\end{figure}

These results on the reversal of the bias field are the basis of
topological phase engineering to create
vortices~\cite{Nakahara2000a,Isoshima2000a,Mottonen2002a,Mottonen2002b}.
In vortex pumping, however, the control sequence is to be cyclic,
and hence we have to turn the spin back to its original position
to end the cycle. If this is carried out with the same
perpendicular field as the first reversal, Eq.~\eqref{rot} implies
that the two reversals cancel each other,
$R_{\hat{\bm{n}}}(-\pi)R_{\hat{\bm{n}}}(\pi)=I$, and the
topologically created vortex unwinds as was observed in
Ref.~\cite{Leanhardt2002a}. In general, only one control parameter
cannot provide geometric effects in cyclic and adiabatic temporal
evolution. Furthermore, since the spin rotation axis in the bias
field reversal does not depend on the absolute strength of the
perpendicular field and the reversal of the perpendicular field
before a subsequent reversal of the bias field results in
$R_{-\hat{\bm{n}}}(-\pi)R_{\hat{\bm{n}}}(\pi)=R_{\hat{\bm{n}}}(2\pi)=I$
for bosons, two different perpendicular fields are required for a
feasible pumping cycle. Such a cycle is shown in Fig.~\ref{fig1}
for our pumping scheme. In fact, the paths between the initial and
the middle point of the cycle can be arbitrary with the constraint
that either~$B_q'$ or~$B_h'$ vanishes always in the adiabatic
evolution.

We begin from the ground state of the system which corresponds to
\begin{equation}
\langle
\bm{r},F,m_z|\Psi_0\rangle=\sqrt{\rho(\bm{r})}\delta_{m_z,-F},
\end{equation}
where~$\rho(\bm{r})$ is the condensate density and~$\delta_{i,j}$
the Kronecker delta~\footnote{Actually, anti-ferromagnetic
atom-atom interactions can render the spin part of the condensate
to deviate from this. Hence, we consider only the ferromagnetic
case, although it is not crucial for the vortex pump to work.}. By
adiabatically ramping up the hexapole field, reversing the bias
field, and ramping down the hexapole field, we have performed the
spin rotation corresponding to Eq.~\eqref{vf}, and achieved the
$4F$-quantum vortex state
$e^{i4F\phi}\sqrt{\rho^{(1)}(\bm{r})}|F,m_z=F\rangle$, where the
superscript denotes possible changes in the particle density due
to the emergence of the vortex. Then, we bring the spin state back
to its initial position by adiabatically ramping up the quadrupole
field, reversing again the bias field, and ramping down the
quadrupole field. This corresponds to the spin
rotation~$R_{\hat{\bm{n}}_q}(-\pi)$ which removes~$2F$ quanta of
vorticity. Thus the total effect of~$n$ adiabatic pumping cycles
on the initial ground state is
\begin{eqnarray}
\langle
\bm{r},F,m_z|\Psi_n\rangle = 
e^{i2nF\phi}\sqrt{\rho^{(2n)}(\bm{r})}\delta_{m_z,-F}.
\end{eqnarray}
We observe that our pump increases the vorticity of the system
by~$2F$ per cycle. Since the particle density in this trapping
scenario is independent of the azimuthal angle~$\phi$, the axial
angular momentum corresponding to this vorticity is $\langle
\Psi_n|\hat{L}_z|\Psi_n\rangle=2nNF\hbar$,
where~$N=\int\rho(\bm{r})\textrm{d}\bm{r}$ is the total particle
number.

The topological phase engineering method has been explained in
terms of accumulation of geometric phases~\cite{Berry1984} for
individual spins of the atoms the condensate is composed
of~\cite{Mottonen2002a,Leanhardt2002a}. This interpretation is
based on the result by Berry~\cite{Berry1984} that an
adiabatically and cyclically turned spin acquires a geometric
phase equal to $-m_F\Omega$, where~$\Omega$ is the solid angle
covered by the path of the spin with the initial
state~$|F,m_F\rangle$. The result of Eq.~\eqref{vf} follows from
the fact that the solid angle covered by the paths of two
different spins separated by an angle~$\theta$ in the~$\phi$
coordinate is~$4\theta$ for the reversal of the bias field in the
presence of the hexapole field. In the case of the vortex pump,
the result follows directly from Berry's argument, since we need
only to consider the path traced by a single spin. In our pumping
scheme, the solid angle~$\Omega$ equals twice the angle between
the normal vectors of the hexapole and the quadrupole fields given
in Eqs.~\eqref{nq} and~\eqref{nh}. This results in the local Berry
phase of~$2\phi$, consistent with pumping two quanta of vorticity
per cycle. Note that the above interpretation of vortex pumping in
terms of the Berry phase is not directly related to the usual
Berry phase accumulated to a macroscopic quantum state of a system
during adiabatic control parameter cycles.

The adiabatic turning of the condensate spin by the magnetic field
requires that the magnetic energy dominates over the kinetic
energy at each spatial point and that the field is rotated slowly
enough to avoid spin flips due to Landau-Zener transitions. Thus
to guarantee adiabaticity, the strength of the external magnetic
field~$|\bm{B}(\bm{r})|$ should be large enough in the condensate
region. However, this condition is not fulfilled
 on the $z$-axis when the bias field crosses zero, and hence it is desirable to avoid particles from entering this area. This can be achieved by introducing an
 additional optical plug potential
\begin{equation}\label{plug}
V_\textrm{plug}(r)=Ae^{-r^2/\delta^2},
\end{equation}
where~$A$ is the strength of the plug and~$\delta$ its width. The
plug can be realized for pancake-shaped condensates by a focused
laser beam as was done, e.g., in Ref.~\cite{Simula2005}. In
addition, to improve adiabaticity during the magnetic field field
reversal, the optical plug tends to stabilize the created
multi-quantum vortex against splitting.

The magnetic trapping potential arises due to the spatial
dependence of the perpendicular field. Thus the magnetic trap
vanishes in the beginning and at the middle point of the pumping
cycle, which also challenges the adiabaticity of the pump. To
prevent the condensate from expanding, we employ an additional
harmonic optical potential
\begin{equation}\label{vopt}
V_\textrm{opt}(r,\phi,z)=\frac{m}{2}(\omega_r^2r^2+\omega_z^2z^2),
\end{equation}
where~$m$ is the atomic mass and~$\omega_k$ the frequency of the
potential along direction~$k$. This kind of potential can be
realized with laser fields~\cite{Pethick2002a}.

We have studied in detail the temporal evolution of a $F=1$
condensate during vortex pumping by modeling it with the
time-dependent Gross-Pitaevskii equation~\cite{Ho1998,Ohmi1998}
\begin{eqnarray}
i\hbar\partial_t\bm{\Psi}(\bm{r})&=&\big[-\frac{\hbar^2}{2m}\nabla^2+V_\textrm{plug}+V_\textrm{opt}+\mu_Bg_F\bm{B}\cdot\bm{\mathcal{F}}
\nonumber \\
&+&
c_0|\bm{\Psi}|^2+c_2\bm{\mathcal{F}}\cdot(\bm{\Psi}^\dagger\bm{\mathcal{F}}\bm{\Psi})\big]\bm{\Psi}(\bm{r}),
\end{eqnarray}
where~$\bm{\Psi}(\bm{r})$ is a vector representation of the spinor
order parameter~$\langle\bm{r}|\Psi\rangle$, $\mu_B$ the Bohr
magneton, $g_F=1/2$
 the Land{$\acute{\textrm{e}}$}~$g$ factor, and $c_0=4\pi\hbar^2(a_0+2a_2)/(3m)$ and $c_2=4\pi\hbar^2(a_2-a_0)/(3m)$ the coupling constants related to $s$-wave scattering lengths~$a_0$ and~$a_2$ of the atoms for different spin channels. We study condensates with repulsive atom-atom interactions, for which the atom-specific constant~$c_0$ is positive.
 The constant~$c_2$ is taken to be negative for the condensate to be ferromagnetic.

 We consider a pancake-shaped condensate for which the~$z$ dependence of
the particle density can be taken to have the Gaussian
form~$e^{-z^2/(2a_z^2)}$, where $a_k=\sqrt{\hbar/(m\omega_k)}$ is
the oscillator length in the direction~$k$. This approximation is
valid for $\omega_z\gg\omega_r$ and implies that the~$z$
dependence of the GP equation can be factored out. We measure the
length in the units of~$a_r$, time in~$1/\omega_r$, energy
in~$\hbar\omega_r$, and magnetic field in~$\hbar\omega_r/\mu_B$,
and normalize the order parameter to unity. This results in a GP
equation which depends only on the dimensionless
parameters~$\tilde{c}_0=2\sqrt{2\pi}N(a_0+2a_2)/(3a_z)$,
$\tilde{c}_ 2=2\sqrt{2\pi}N(a_2-a_0)/(3a_z)$, $A/(\hbar\omega_r)$,
and $\delta/a_r$, together with the time-dependent magnetic field.

\begin{figure}[h]
    \begin{center}
   \includegraphics[width=.45\textwidth]{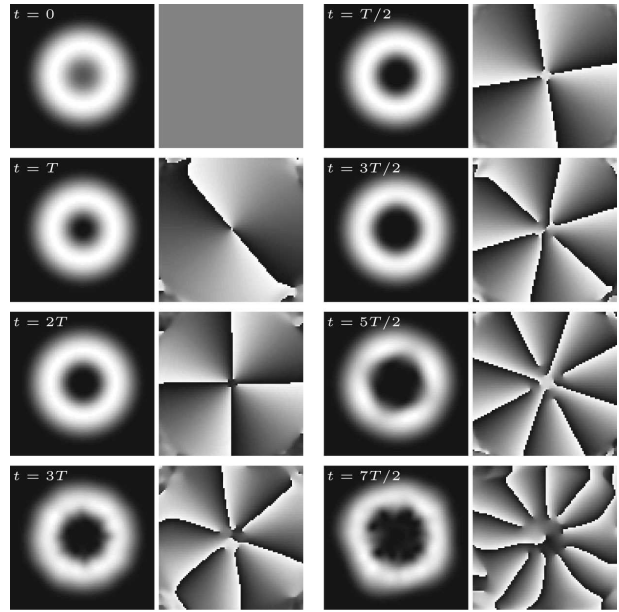}
    \end{center}
   \caption{\label{fig2} Particle density and the complex phase of the $|F=1,m_z=-1\rangle$ (left columns) and $|F=1,m_z=1\rangle$ (right columns) components of the order parameter at half integer multiples of the pumping period~$T=360/\omega_r$. The field of view is $12a_r\times 12a_r$.}
\end{figure}

We choose the parameters used in our numerical computations
according to $^{87}$Rb~\cite{Pethick2002a}, such that
$c_2/c_0=-0.01$. Furthermore, we set $\tilde{c}_0=250$. We note
that the atom-atom interactions do not play a significant role in
the operation of the vortex pump and qualitatively similar results
are obtained, e.g., in the non-interacting case and for
anti-ferromagnetic interactions.
The parameters of the optical plug defined in Eq.~\eqref{plug} are
chosen as $A=10\hbar\omega_r$ and $\delta=2a_r$. The maximum
values of the magnetic fields are taken to be $B_z^\textrm{max}=40
\hbar\omega_r/\mu_B$, $B_q'{^\textrm{max}}=
\hbar\omega_r/(\mu_Ba_r)$, and $B_h'{^\textrm{max}}=
\hbar\omega_r/(\mu_Ba_r)$.

\begin{figure}[h]
    \begin{center}
   \includegraphics[width=.47\textwidth]{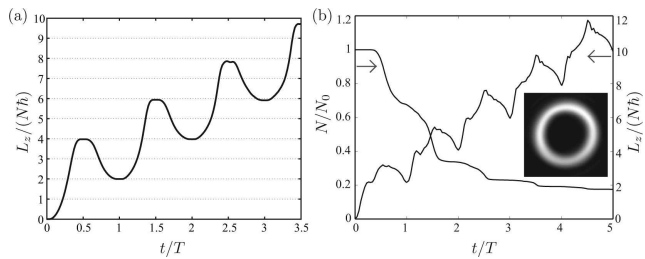}
    \end{center}
   \caption{\label{fig3} Accumulation of axial angular momentum to the system during (a) adiabatic vortex pumping ($T=360/\omega_r$) and (b) pumping without any optical potentials ($T=10/\omega_r$), for which the fraction of atoms in the magnetic trap decreases in time as shown. The inset shows the particle density after five full cycles of non-adiabatic pumping with the field of view $12a_r\times 12a_r$.}
\end{figure}

Before beginning the pumping cycle, the ground state of the
condensate is found by a relaxation method with the magnetic field
parameters set to values corresponding to $t=0$ in
Fig.~\ref{fig1}. Figure~\ref{fig2} shows the particle density and
the phase of the relevant components of the condensate during
vortex pumping. The accumulation of two quanta of vorticity per
cycle is clearly visible in the complex phase of the condensate.
As more vortices are pumped to the BEC, the density depletion
along the $z$-axis grows due to increasing core size of the vortex
with increasing vorticity. The energetic and dynamic stability of
the multi-quantum vortex is not guaranteed if the core area is
much larger than the size of the optical plug, and hence
initiation of splitting of the multi-quantum vortex into
single-quantum vortices is observed for the ten-quantum vortex
shown in Fig.~\ref{fig2}. The observed instability can be overcome
by increasing the strength and the width of the plug.

The number of vortices in each spin component is quantized, but
the axial angular momentum is a continuous quantity as shown in
Fig.~\ref{fig3}(a). The figure illustrates the increment of
vorticity by four during the field reversal in the presence of the
hexapole field and the decrement of it by two when the field is
brought back to its initial position with the quadrupole field
turned on. The deviation of the axial angular momentum
from~$2n\hbar N$ after~$n$ pumping cycles is due to excitations
away from the instantaneous eigenstate the pump is operated in.

Although the optical potentials given by Eqs.~\eqref{plug}
and~\eqref{vopt} are experimentally realizable using standard
techniques, they can render the experiment rather complicated.
However, the vortex pump can also be operated non-adiabatically
without any optical potentials, since the spin components which do
not follow the magnetic field are not trapped, and hence leave the
condensate region. In fact, no optical fields were employed in the
experiments of Ref.~\cite{Leanhardt2002a} with the cost of loosing
roughly half of the atoms during the reversal of the bias field.
Furthermore, if the maximum value of the bias field is much larger
than that of the perpendicular field, an instantaneous switch
between the quadrupole and hexapole field results in a negligible
atom loss. Thus we suggest that the first experimental realization
of the vortex pump could be in a set-up corresponding to the one
used in Refs.~\cite{Leanhardt2002a,Leanhardt2003a,Shin2004a} but
with three additional coils to generate the hexapole field. Our
numerical studies on vortex pumping with this technique show that
in fact the total atom loss is considerably suppressed by the
presence of the pumped multi-quantum vortex along the $z$-axis,
causing the particle density to vanish there, see
Fig.~\ref{fig3}(b). Since the cycle can be traversed much faster
than in the fully adiabatic case, there is no time for the
multi-quantum vortex to split, and hence a clean ten-quantum
vortex is observed in the inset of Fig.~\ref{fig3}(b) after five
pumping cycles.

In conclusion, we have shown how to implement a vortex pump for
spinor Bose-Einstein condensates using experimentally feasible
techniques. Adiabatic and non-adiabatic operation of the pump was
demonstrated computationally in the case of hyperfine spin $F=1$
by pumping vortices up to twelve circulation quanta to the
condensate. In contrast to the experimentally created giant
vortex~\cite{Engels2003}, our scheme produces a pure multi-quantum
vortex without accompanying single-quantum vortices. This vortex
pump is an interesting example of adiabatic quantum dynamics for
which the control parameters of the system are varied cyclically
but the system does not return to its initial eigenspace.
Our vortex pump paves the way for vortex-based applications of
BECs since it can work as a vortex source, similar to a spin
polarized current source in spintronics.

We acknowledge the Academy of Finland, Finnish Cultural
Foundation, Wihuri foundation, and V\"ais\"al\"a foundation for
financial support.

\bibliography{manu}

\begin{thebibliography}{25}
\expandafter\ifx\csname natexlab\endcsname\relax\def\natexlab#1{#1}\fi
\expandafter\ifx\csname bibnamefont\endcsname\relax
  \def\bibnamefont#1{#1}\fi
\expandafter\ifx\csname bibfnamefont\endcsname\relax
  \def\bibfnamefont#1{#1}\fi
\expandafter\ifx\csname citenamefont\endcsname\relax
  \def\citenamefont#1{#1}\fi
\expandafter\ifx\csname url\endcsname\relax
  \def\url#1{\texttt{#1}}\fi
\expandafter\ifx\csname urlprefix\endcsname\relax\def\urlprefix{URL }\fi
\providecommand{\bibinfo}[2]{#2}
\providecommand{\eprint}[2][]{\url{#2}}

\bibitem[{\citenamefont{Pethick and Smith}(2002)}]{Pethick2002a}
\bibinfo{author}{\bibfnamefont{C.~J.} \bibnamefont{Pethick}} \bibnamefont{and}
  \bibinfo{author}{\bibfnamefont{H.}~\bibnamefont{Smith}},
  \emph{\bibinfo{title}{{B}ose-{E}instein Condensation in Dilute Gases}}
  (\bibinfo{publisher}{Cambridge University Press},
  \bibinfo{address}{Cambridge}, \bibinfo{year}{2002}).

\bibitem[{\citenamefont{Matthews et~al.}(1999)\citenamefont{Matthews, Anderson,
  Haljan, Hall, Wieman, and Cornell}}]{Matthews1999b}
\bibinfo{author}{\bibfnamefont{M.~R.} \bibnamefont{Matthews}},
  \bibinfo{author}{\bibfnamefont{B.~P.} \bibnamefont{Anderson}},
  \bibinfo{author}{\bibfnamefont{P.~C.} \bibnamefont{Haljan}},
  \bibinfo{author}{\bibfnamefont{D.~S.} \bibnamefont{Hall}},
  \bibinfo{author}{\bibfnamefont{C.~E.} \bibnamefont{Wieman}},
  \bibnamefont{and} \bibinfo{author}{\bibfnamefont{E.~A.}
  \bibnamefont{Cornell}}, \bibinfo{journal}{Phys. Rev. Lett.}
  \textbf{\bibinfo{volume}{83}}, \bibinfo{pages}{2498} (\bibinfo{year}{1999}).

\bibitem[{\citenamefont{Williams and Holland}(1999)}]{Williams1999c}
\bibinfo{author}{\bibfnamefont{J.~E.} \bibnamefont{Williams}} \bibnamefont{and}
  \bibinfo{author}{\bibfnamefont{M.~J.} \bibnamefont{Holland}},
  \bibinfo{journal}{Nature} \textbf{\bibinfo{volume}{401}},
  \bibinfo{pages}{568} (\bibinfo{year}{1999}).

\bibitem[{\citenamefont{Inouye et~al.}(2001)\citenamefont{Inouye, Gupta,
  Rosenband, Chikkatur, G{\"o}rlitz, Gustavson, Leanhardt, Pritchard, and
  Ketterle}}]{Inouye2001a}
\bibinfo{author}{\bibfnamefont{S.}~\bibnamefont{Inouye}},
  \bibinfo{author}{\bibfnamefont{S.}~\bibnamefont{Gupta}},
  \bibinfo{author}{\bibfnamefont{T.}~\bibnamefont{Rosenband}},
  \bibinfo{author}{\bibfnamefont{A.~P.} \bibnamefont{Chikkatur}},
  \bibinfo{author}{\bibfnamefont{A.}~\bibnamefont{G{\"o}rlitz}},
  \bibinfo{author}{\bibfnamefont{T.~L.} \bibnamefont{Gustavson}},
  \bibinfo{author}{\bibfnamefont{A.~E.} \bibnamefont{Leanhardt}},
  \bibinfo{author}{\bibfnamefont{D.~E.} \bibnamefont{Pritchard}},
  \bibnamefont{and} \bibinfo{author}{\bibfnamefont{W.}~\bibnamefont{Ketterle}},
  \bibinfo{journal}{Phys. Rev. Lett.} \textbf{\bibinfo{volume}{87}},
  \bibinfo{pages}{080402} (\bibinfo{year}{2001}).

\bibitem[{\citenamefont{Scherer et~al.}(2007)\citenamefont{Scherer, Weiler,
  Neely, and Anderson}}]{Scherer2007}
\bibinfo{author}{\bibfnamefont{D.~R.} \bibnamefont{Scherer}},
  \bibinfo{author}{\bibfnamefont{C.~N.} \bibnamefont{Weiler}},
  \bibinfo{author}{\bibfnamefont{T.~W.} \bibnamefont{Neely}}, \bibnamefont{and}
  \bibinfo{author}{\bibfnamefont{B.~P.} \bibnamefont{Anderson}},
  \bibinfo{journal}{Phys. Rev. Lett.} \textbf{\bibinfo{volume}{98}},
  \bibinfo{pages}{110402} (\bibinfo{year}{2007}).

\bibitem[{\citenamefont{Madison et~al.}(2000)\citenamefont{Madison, Chevy,
  Wohlleben, and Dalibard}}]{Madison2000a}
\bibinfo{author}{\bibfnamefont{K.~W.} \bibnamefont{Madison}},
  \bibinfo{author}{\bibfnamefont{F.}~\bibnamefont{Chevy}},
  \bibinfo{author}{\bibfnamefont{W.}~\bibnamefont{Wohlleben}},
  \bibnamefont{and} \bibinfo{author}{\bibfnamefont{J.}~\bibnamefont{Dalibard}},
  \bibinfo{journal}{Phys. Rev. Lett.} \textbf{\bibinfo{volume}{84}},
  \bibinfo{pages}{806} (\bibinfo{year}{2000}).

\bibitem[{\citenamefont{Hodby et~al.}(2001)\citenamefont{Hodby, Hechenblaikner,
  Hopkins, Marag\`o, and Foot}}]{Hodby2001}
\bibinfo{author}{\bibfnamefont{E.}~\bibnamefont{Hodby}},
  \bibinfo{author}{\bibfnamefont{G.}~\bibnamefont{Hechenblaikner}},
  \bibinfo{author}{\bibfnamefont{S.~A.} \bibnamefont{Hopkins}},
  \bibinfo{author}{\bibfnamefont{O.~M.} \bibnamefont{Marag\`o}},
  \bibnamefont{and} \bibinfo{author}{\bibfnamefont{C.~J.} \bibnamefont{Foot}},
  \bibinfo{journal}{Phys. Rev. Lett.} \textbf{\bibinfo{volume}{88}},
  \bibinfo{pages}{010405} (\bibinfo{year}{2001}).

\bibitem[{\citenamefont{{M{\" o}tt{\" o}nen} et~al.}(2003)\citenamefont{{M{\"
  o}tt{\" o}nen}, {Mizushima}, {Isoshima}, {Salomaa}, and
  {Machida}}}]{Mottonen2003a}
\bibinfo{author}{\bibfnamefont{M.}~\bibnamefont{{M{\" o}tt{\" o}nen}}},
  \bibinfo{author}{\bibfnamefont{T.}~\bibnamefont{{Mizushima}}},
  \bibinfo{author}{\bibfnamefont{T.}~\bibnamefont{{Isoshima}}},
  \bibinfo{author}{\bibfnamefont{M.~M.} \bibnamefont{{Salomaa}}},
  \bibnamefont{and}
  \bibinfo{author}{\bibfnamefont{K.}~\bibnamefont{{Machida}}},
  \bibinfo{journal}{Phys. Rev. A} \textbf{\bibinfo{volume}{68}},
  \bibinfo{pages}{023611} (\bibinfo{year}{2003}).

\bibitem[{\citenamefont{Shin et~al.}(2004)\citenamefont{Shin, Saba,
  Vengalattore, Pasquini, Sanner, Leanhardt, Prentiss, Pritchard, and
  Ketterle}}]{Shin2004a}
\bibinfo{author}{\bibfnamefont{Y.}~\bibnamefont{Shin}},
  \bibinfo{author}{\bibfnamefont{M.}~\bibnamefont{Saba}},
  \bibinfo{author}{\bibfnamefont{M.}~\bibnamefont{Vengalattore}},
  \bibinfo{author}{\bibfnamefont{T.~A.} \bibnamefont{Pasquini}},
  \bibinfo{author}{\bibfnamefont{C.}~\bibnamefont{Sanner}},
  \bibinfo{author}{\bibfnamefont{A.~E.} \bibnamefont{Leanhardt}},
  \bibinfo{author}{\bibfnamefont{M.}~\bibnamefont{Prentiss}},
  \bibinfo{author}{\bibfnamefont{D.~E.} \bibnamefont{Pritchard}},
  \bibnamefont{and} \bibinfo{author}{\bibfnamefont{W.}~\bibnamefont{Ketterle}},
  \bibinfo{journal}{Phys. Rev. Lett.} \textbf{\bibinfo{volume}{93}},
  \bibinfo{pages}{160406} (\bibinfo{year}{2004}).

\bibitem[{\citenamefont{Huhtam\"aki et~al.}(2006)\citenamefont{Huhtam\"aki,
  M\"ott\"onen, Isoshima, Pietil\"a, and Virtanen}}]{Huhtamaki2006a}
\bibinfo{author}{\bibfnamefont{J.~A.~M.} \bibnamefont{Huhtam\"aki}},
  \bibinfo{author}{\bibfnamefont{M.}~\bibnamefont{M\"ott\"onen}},
  \bibinfo{author}{\bibfnamefont{T.}~\bibnamefont{Isoshima}},
  \bibinfo{author}{\bibfnamefont{V.}~\bibnamefont{Pietil\"a}},
  \bibnamefont{and} \bibinfo{author}{\bibfnamefont{S.~M.~M.}
  \bibnamefont{Virtanen}}, \bibinfo{journal}{Phys. Rev. Lett.}
  \textbf{\bibinfo{volume}{97}}, \bibinfo{pages}{110406}
  (\bibinfo{year}{2006}).

\bibitem[{\citenamefont{Abo-Shaeer et~al.}(2001)\citenamefont{Abo-Shaeer,
  Raman, Vogels, and Ketterle}}]{Abo-Shaeer2001a}
\bibinfo{author}{\bibfnamefont{J.~R.} \bibnamefont{Abo-Shaeer}},
  \bibinfo{author}{\bibfnamefont{C.}~\bibnamefont{Raman}},
  \bibinfo{author}{\bibfnamefont{J.~M.} \bibnamefont{Vogels}},
  \bibnamefont{and} \bibinfo{author}{\bibfnamefont{W.}~\bibnamefont{Ketterle}},
  \bibinfo{journal}{Science} \textbf{\bibinfo{volume}{292}},
  \bibinfo{pages}{476} (\bibinfo{year}{2001}).

\bibitem[{\citenamefont{Raman et~al.}(2001)\citenamefont{Raman, Abo-Shaeer,
  Vogels, Xu, and Ketterle}}]{Raman2001a}
\bibinfo{author}{\bibfnamefont{C.}~\bibnamefont{Raman}},
  \bibinfo{author}{\bibfnamefont{J.~R.} \bibnamefont{Abo-Shaeer}},
  \bibinfo{author}{\bibfnamefont{J.~M.} \bibnamefont{Vogels}},
  \bibinfo{author}{\bibfnamefont{K.}~\bibnamefont{Xu}}, \bibnamefont{and}
  \bibinfo{author}{\bibfnamefont{W.}~\bibnamefont{Ketterle}},
  \bibinfo{journal}{Phys. Rev. Lett.} \textbf{\bibinfo{volume}{87}},
  \bibinfo{pages}{210402} (\bibinfo{year}{2001}).

\bibitem[{\citenamefont{Andersen et~al.}(2006)\citenamefont{Andersen, Ryu,
  Clad\'e, Natarajan, Vaziri, Helmerson, and Phillips}}]{Andersen2006}
\bibinfo{author}{\bibfnamefont{M.~F.} \bibnamefont{Andersen}},
  \bibinfo{author}{\bibfnamefont{C.}~\bibnamefont{Ryu}},
  \bibinfo{author}{\bibfnamefont{P.}~\bibnamefont{Clad\'e}},
  \bibinfo{author}{\bibfnamefont{V.}~\bibnamefont{Natarajan}},
  \bibinfo{author}{\bibfnamefont{A.}~\bibnamefont{Vaziri}},
  \bibinfo{author}{\bibfnamefont{K.}~\bibnamefont{Helmerson}},
  \bibnamefont{and} \bibinfo{author}{\bibfnamefont{W.~D.}
  \bibnamefont{Phillips}}, \bibinfo{journal}{Phys. Rev. Lett.}
  \textbf{\bibinfo{volume}{97}}, \bibinfo{pages}{170406}
  (\bibinfo{year}{2006}).

\bibitem[{\citenamefont{Engels et~al.}(2003)\citenamefont{Engels, Coddington,
  Haljan, Schweikhard, and Cornell}}]{Engels2003}
\bibinfo{author}{\bibfnamefont{P.}~\bibnamefont{Engels}},
  \bibinfo{author}{\bibfnamefont{I.}~\bibnamefont{Coddington}},
  \bibinfo{author}{\bibfnamefont{P.~C.} \bibnamefont{Haljan}},
  \bibinfo{author}{\bibfnamefont{V.}~\bibnamefont{Schweikhard}},
  \bibnamefont{and} \bibinfo{author}{\bibfnamefont{E.~A.}
  \bibnamefont{Cornell}}, \bibinfo{journal}{Phys. Rev. Lett.}
  \textbf{\bibinfo{volume}{90}}, \bibinfo{pages}{170405}
  (\bibinfo{year}{2003}).

\bibitem[{\citenamefont{Nakahara et~al.}(2000)\citenamefont{Nakahara, Isoshima,
  Machida, Ogawa, and Ohmi}}]{Nakahara2000a}
\bibinfo{author}{\bibfnamefont{M.}~\bibnamefont{Nakahara}},
  \bibinfo{author}{\bibfnamefont{T.}~\bibnamefont{Isoshima}},
  \bibinfo{author}{\bibfnamefont{K.}~\bibnamefont{Machida}},
  \bibinfo{author}{\bibfnamefont{S.-I.} \bibnamefont{Ogawa}}, \bibnamefont{and}
  \bibinfo{author}{\bibfnamefont{T.}~\bibnamefont{Ohmi}},
  \bibinfo{journal}{Physica B} \textbf{\bibinfo{volume}{284--288}},
  \bibinfo{pages}{17} (\bibinfo{year}{2000}).

\bibitem[{\citenamefont{Isoshima et~al.}(2000)\citenamefont{Isoshima, Nakahara,
  Ohmi, and Machida}}]{Isoshima2000a}
\bibinfo{author}{\bibfnamefont{T.}~\bibnamefont{Isoshima}},
  \bibinfo{author}{\bibfnamefont{M.}~\bibnamefont{Nakahara}},
  \bibinfo{author}{\bibfnamefont{T.}~\bibnamefont{Ohmi}}, \bibnamefont{and}
  \bibinfo{author}{\bibfnamefont{K.}~\bibnamefont{Machida}},
  \bibinfo{journal}{Phys. Rev. A} \textbf{\bibinfo{volume}{61}},
  \bibinfo{pages}{063610} (\bibinfo{year}{2000}).

\bibitem[{\citenamefont{Ogawa et~al.}(2002)\citenamefont{Ogawa, M\"ott\"onen,
  Nakahara, Ohmi, and Shimada}}]{Mottonen2002a}
\bibinfo{author}{\bibfnamefont{S.-I.} \bibnamefont{Ogawa}},
  \bibinfo{author}{\bibfnamefont{M.}~\bibnamefont{M\"ott\"onen}},
  \bibinfo{author}{\bibfnamefont{M.}~\bibnamefont{Nakahara}},
  \bibinfo{author}{\bibfnamefont{T.}~\bibnamefont{Ohmi}}, \bibnamefont{and}
  \bibinfo{author}{\bibfnamefont{H.}~\bibnamefont{Shimada}},
  \bibinfo{journal}{Phys. Rev. A} \textbf{\bibinfo{volume}{66}},
  \bibinfo{pages}{013617} (\bibinfo{year}{2002}).

\bibitem[{\citenamefont{M\"ott\"onen et~al.}(2002)\citenamefont{M\"ott\"onen,
  Matsumoto, Nakahara, and Ohmi}}]{Mottonen2002b}
\bibinfo{author}{\bibfnamefont{M.}~\bibnamefont{M\"ott\"onen}},
  \bibinfo{author}{\bibfnamefont{N.}~\bibnamefont{Matsumoto}},
  \bibinfo{author}{\bibfnamefont{M.}~\bibnamefont{Nakahara}}, \bibnamefont{and}
  \bibinfo{author}{\bibfnamefont{T.}~\bibnamefont{Ohmi}}, \bibinfo{journal}{J.
  Phys.: Condens. Matter} \textbf{\bibinfo{volume}{14}}, \bibinfo{pages}{13481}
  (\bibinfo{year}{2002}).

\bibitem[{\citenamefont{Leanhardt et~al.}(2002)\citenamefont{Leanhardt,
  G{\"o}rlitz, Chikkatur, Kielpinski, Shin, Pritchard, and
  Ketterle}}]{Leanhardt2002a}
\bibinfo{author}{\bibfnamefont{A.~E.} \bibnamefont{Leanhardt}},
  \bibinfo{author}{\bibfnamefont{A.}~\bibnamefont{G{\"o}rlitz}},
  \bibinfo{author}{\bibfnamefont{A.~P.} \bibnamefont{Chikkatur}},
  \bibinfo{author}{\bibfnamefont{D.}~\bibnamefont{Kielpinski}},
  \bibinfo{author}{\bibfnamefont{Y.}~\bibnamefont{Shin}},
  \bibinfo{author}{\bibfnamefont{D.~E.} \bibnamefont{Pritchard}},
  \bibnamefont{and} \bibinfo{author}{\bibfnamefont{W.}~\bibnamefont{Ketterle}},
  \bibinfo{journal}{Phys. Rev. Lett.} \textbf{\bibinfo{volume}{89}},
  \bibinfo{pages}{190403} (\bibinfo{year}{2002}).

\bibitem[{\citenamefont{Cooper et~al.}(2001)\citenamefont{Cooper, Wilkin, and
  Gunn}}]{Cooper2001}
\bibinfo{author}{\bibfnamefont{N.~R.} \bibnamefont{Cooper}},
  \bibinfo{author}{\bibfnamefont{N.~K.} \bibnamefont{Wilkin}},
  \bibnamefont{and} \bibinfo{author}{\bibfnamefont{J.~M.~F.}
  \bibnamefont{Gunn}}, \bibinfo{journal}{Phys. Rev. Lett.}
  \textbf{\bibinfo{volume}{87}}, \bibinfo{pages}{120405}
  (\bibinfo{year}{2001}).

\bibitem[{\citenamefont{Berry}(1984)}]{Berry1984}
\bibinfo{author}{\bibfnamefont{M.~V.} \bibnamefont{Berry}},
  \bibinfo{journal}{Proc. R. Soc. London, Ser. A}
  \textbf{\bibinfo{volume}{392}}, \bibinfo{pages}{45} (\bibinfo{year}{1984}).

\bibitem[{\citenamefont{Simula et~al.}(2005)\citenamefont{Simula, Engels,
  Coddington, Schweikhard, Cornell, and Ballagh}}]{Simula2005}
\bibinfo{author}{\bibfnamefont{T.~P.} \bibnamefont{Simula}},
  \bibinfo{author}{\bibfnamefont{P.}~\bibnamefont{Engels}},
  \bibinfo{author}{\bibfnamefont{I.}~\bibnamefont{Coddington}},
  \bibinfo{author}{\bibfnamefont{V.}~\bibnamefont{Schweikhard}},
  \bibinfo{author}{\bibfnamefont{E.~A.} \bibnamefont{Cornell}},
  \bibnamefont{and} \bibinfo{author}{\bibfnamefont{R.~J.}
  \bibnamefont{Ballagh}}, \bibinfo{journal}{Phys. Rev. Lett.}
  \textbf{\bibinfo{volume}{94}}, \bibinfo{pages}{080404}
  (\bibinfo{year}{2005}).

\bibitem[{\citenamefont{Ho}(1998)}]{Ho1998}
\bibinfo{author}{\bibfnamefont{T.-L.} \bibnamefont{Ho}},
  \bibinfo{journal}{Phys. Rev. Lett.} \textbf{\bibinfo{volume}{81}},
  \bibinfo{pages}{742} (\bibinfo{year}{1998}).

\bibitem[{\citenamefont{Ohmi and Machida}(1998)}]{Ohmi1998}
\bibinfo{author}{\bibfnamefont{T.}~\bibnamefont{Ohmi}} \bibnamefont{and}
  \bibinfo{author}{\bibfnamefont{K.}~\bibnamefont{Machida}},
  \bibinfo{journal}{J. Phys. Soc. Jpn.} \textbf{\bibinfo{volume}{67}},
  \bibinfo{pages}{1822} (\bibinfo{year}{1998}).

\bibitem[{\citenamefont{Leanhardt et~al.}(2003)\citenamefont{Leanhardt, Shin,
  Kielpinski, Pritchard, and Ketterle}}]{Leanhardt2003a}
\bibinfo{author}{\bibfnamefont{A.~E.} \bibnamefont{Leanhardt}},
  \bibinfo{author}{\bibfnamefont{Y.}~\bibnamefont{Shin}},
  \bibinfo{author}{\bibfnamefont{D.}~\bibnamefont{Kielpinski}},
  \bibinfo{author}{\bibfnamefont{D.~E.} \bibnamefont{Pritchard}},
  \bibnamefont{and} \bibinfo{author}{\bibfnamefont{W.}~\bibnamefont{Ketterle}},
  \bibinfo{journal}{Phys. Rev. Lett.} \textbf{\bibinfo{volume}{90}},
  \bibinfo{pages}{140403} (\bibinfo{year}{2003}).

\end{thebibliography}

\end{document}